\documentclass[12pt]{article} 
\title{Massive Particle Fields, with Momentum Matrices}  
\author{{\it Richard Shurtleff~}\thanks{affiliation and mailing 
address: Department of Applied Mathematics and Sciences, 
Wentworth Institute of Technology, 550 Huntington Avenue, 
Boston, MA, USA, ZIP 02115, telephone number: (617) 989-4338, fax 
number: (617) 989-4591 , e-mail address: shurtleffr@wit.edu}} 
\begin{document} 
          
\maketitle 

\begin{abstract} 

It is well known that an essentially unique free massive particle field of given spin can be constructed as a linear combination of the annihilation and creation operators for free single particle states. And the given spin determines the transformation of the field components for rotations and boosts. But the usual construction restricts translations; the components are simply invariant under translations. This paper generalizes the construction so that translations may have a nontrivial effect on field components.

PACS:  11.30.Cp 	Lorentz and Poincare invariance

\end{abstract}

%\pagebreak

\section{Introduction}

Successive infinitesimal rotations, boosts, and translations transform spacetime yet preserve the spacetime metric. The inhomogeneous Lorentz group, or Poincar\'{e} group, of such transformations has both unitary and nonunitary representations. 

By basic quantum theory, single particle states and their annihilation and creation operators that remove or add a single particle state to a multiparticle state must transform by unitary representations (reps) of the Poincar\'{e} group. Since momentum generators commute, the single particle states are sums of eigenstates of momentum. Then the annihilation and creation operators that remove or add momentum eigenstates are defined over the allowed values of particle momenta. 

Among the quantities in physics that transform with nonunitary representations of the Poincar\'{e} group are fields, multicomponent objects defined over spacetime events. It is remarkable that fields can be constructed that are linear combinations of operators and with invariant coefficients$,^{\cite{fields1,fields2}}$ sometimes called the Invariant Coefficient Hypothesis. 

The hypothesis yields unique free massive particle fields, within normalization factors. However the usual derivation limits the transformations of the field components to nonunitary reps of rotations and boosts, neglecting translations. 

In this paper it is shown that it is mathematically unnecessary to limit component transformations to rotations and boosts. We show that the usual derivation can be modified so that the field components transform with nonunitary reps of rotations, boosts and translations, the full complement of spacetime symmetries connected to the identity.

The result is that the field is obtained by applying a pure translation through a displacement $x$ to the translation-deficient field obtained conventionally. 
Let $D(1,x)$ indicate a translation by displacement $x,$ then the field $\psi$ found here can be written as 
\begin{equation} \label{intro3} \psi_{l}(x) = D_{l \bar{l}}(1,x) \psi_{\bar{l}}^{(0)}(x) \, ,
\end{equation}
where $\psi^{(0)}$ is the field found with the restriction that the transformations of the field components are limited to rotations and boosts. The massless class has been shown in an earlier paper$^{\cite{S2}}$ to have the same translation matrix dependence. Thus, for both the massive class and the massless class, the effect of allowing nontrivial translations is to multiply the translation-deficient field by the translation matrix for the displacement $x.$ 

Translations are generated by the momentum and the momentum generators have a free scale factor because the Poincar\'{e} algebra is homogeneous in momentum generators. Predictions would require an investigation of the quantum mechanics of the generalized fields.

Section \ref{derive} follows the procedure from Weinberg$^{ \cite{fields1}}$ closely, except that the field $\psi$ is required to transform with the Poincar\'{e} group that includes translations. The result is described above with equation (\ref{intro3}). A brief discussion is presented in Section \ref{discuss}. Appendix \ref{B} provides a set of exercises and problems to develop understanding and instigate further study.

\section{Massive Fields } \label{derive}

The calculation here follows the approach explained in detail by Weinberg$.^{ \cite{fields1}}$

Translation is a spacetime symmetry whose generators, the momentum operators, commute. Thus, by basic quantum theory, single particle states can be expressed in terms of momentum eigenstates whose eigenvalues are the `momentum'  $p.$ Let $\Psi_{p,\sigma}$ be a single particle eigenstate of momentum,
\begin{equation} \label{eigen}
     P^{\mu}_{(\Psi)} \Psi_{p,\sigma} = p^{\mu} \Psi_{p,\sigma} \, ,
\end{equation}
where $P_{(\Psi)}$ is the momentum operator for the representation of the Poincar\'{e} group realized by the spacetime symmetries of the states $\Psi_{p,\sigma}$ and $\sigma$ is the $z$-component of spin. The annihilation and creation operators $a_{\sigma}({\overrightarrow{p}})$ and $a^{\dagger}_{\sigma}({\overrightarrow{p}})$ remove or add a single particle state with momentum $p$ and spin component $\sigma$ to any multiparticle state. 

The annihilation and creation operators are defined over the appropriate Wigner class of momenta$,^{\cite{Wigner1,W1}}$ $p^2$ = $-m^2$ with $p^{t} \geq m > 0$ and $m$ is the particle mass. The energy $p^{t}$ can be determined from the spatial three-vector $\overrightarrow{p} ,$ $p^{t}$ = $(m^2+\overrightarrow{p}^2)^{1/2},$ so the operators depend on the three spatial components of $\overrightarrow{p}.$ 

The field $\psi_{l}(x)$ is constructed as a linear combination of annihilation and creation operators, $\psi_{l}(x)$ = $\kappa \psi^{+}_{l}(x)$ + $\lambda \psi^{-}_{l}(x),$ where the annihilation field $\psi^{+}$ and the creation field $\psi^{-}$ are given by 
%\begin{equation} \label{psi+}
$$\psi^{+}_{l}(x) = \sum_{\sigma} \int d^3 p \enspace u_{l\sigma}(x,{\overrightarrow{p}}) a_{\sigma}({\overrightarrow{p}})  \, ,$$
%\end{equation}
\begin{equation} \label{psi+}
\psi^{-}_{l}(x) = \sum_{\sigma} \int d^3 p \enspace v_{l\sigma}(x,{\overrightarrow{p}}) a^{\dagger}_{\sigma}({\overrightarrow{p}})  \, .
\end{equation}
The coefficients $u$ and $v$ depend on both $x$ and${\overrightarrow{p}}$ and they connect the field components with the operator components. Since the fields are linear combinations of operators, a field is also an operator. We reserve the term `operator' for the annihilation and creation operators. 

The coefficients $u$ and $v$ can be determined from the ways the fields and operators transform to preserve spacetime symmetries: ({\it{i}}) the operators transform under Poincar\'{e} transformations with a unitary representation (rep), ({\it{ii}}) the coefficients are required to be invariant under Poincar\'{e} transformations and ({\it{iii}}) the field transforms by a nonunitary rep of Poincar\'{e} transformations. Remarkably, the unitary transformations of the operators $a$ and $a^{\dagger}$ can produce a non-unitary transformation of the fields $\psi^{\pm}_{l}$. The three transformation regimens constrain the coefficients so much that the coefficients are essentially determined. 

The annihilation and creation operators transform by a unitary representation of the Poincar\'{e} transformation $(\Lambda,b).^{\cite{fields1,fields2}}$ One has
$$U(\Lambda,b) a_{\sigma}({\overrightarrow{p}}) {U}^{-1}(\Lambda,b) = e^{i \Lambda p \cdot \epsilon(\Lambda,b)} \sqrt{\frac{(\Lambda p)^t}{p^t}}  \sum_{\bar{\sigma}} D^{(j)}_{\sigma \bar{\sigma}}(W^{-1})  a_{\bar{\sigma}}({\overrightarrow{\Lambda p}}) \, ,$$
\begin{equation} \label{Da+}
U(\Lambda,b) a^{\dagger}_{\sigma}({\overrightarrow{p}}) {U}^{-1}(\Lambda,b) = e^{-i \Lambda p \cdot \epsilon(\Lambda,b)} \sqrt{\frac{(\Lambda p)^t}{p^t}}  \sum_{\bar{\sigma}} D^{(j)\ast}_{\sigma \bar{\sigma}}(W^{-1})  a^{\dagger}_{\bar{\sigma}}({\overrightarrow{\Lambda p}}) \, ,
\end{equation}
where $j$ is the spin of the particle and the transformation $W$ is a rotation given by
\begin{equation} \label{WLp}
W(\Lambda, p) = L^{-1}(\Lambda p)\Lambda L(p) \, ,
\end{equation}
with $L(p) $ a standard transformation taking $k^{\mu}$ = $\{0,0,0,m\}$ to $p.$ $L(p) $ rotates $\overrightarrow{p}$ to the $z$ direction, followed by a boost along $z$, followed by the rotation back to $\overrightarrow{p}.$ Since $W(\Lambda, p)k$ = $k,$ it follows that $W$ is a rotation. The matrix $D^{(j)}_{\sigma \bar{\sigma}}(W^{-1}(\Lambda,p))$ represents the rotation $W^{-1}$ for spin $j.$

The unitary transformation $U(\Lambda,b)$ is required to have the effect of a nonunitary transformation on the fields. One requires that 
\begin{equation} \label{Dpsi}
U(\Lambda,b) \psi^{\pm}_{l}(x) {U}^{-1}(\Lambda,b) = \sum_{\bar{l}} D^{-1}_{l \bar{l}}(\Lambda,b)  \psi^{\pm}_{\bar{l}}(\Lambda x + b) \, ,
\end{equation}
where $\Lambda x + b$ are the transformed coordinates of the given event and $D(\Lambda,b)$ is the nonunitary matrix representing the spacetime transformation $(\Lambda,b).$ The rep $D(\Lambda,b)$ is determined by the spin composition of $\psi,$ within equivalences. Let the spin composition of the field $\psi$ be  $(A,B)\oplus(C,D)\oplus \ldots \,.$ The standard angular momentum generators and momentum generators for the nonunitary rep $D(\Lambda,b)$ can be found in Ref. \cite{S}.

Recall that the way a multicomponent object transforms with rotations and boosts determines its spin composition. The spacetime symmetries of the Poincar\'{e} group include also translations and that implies that a spin $(A,B)$ field must be combined with one or more fields of allowed spins $(A\pm 1/2,B\pm 1/2)$ in order to have nontrivial translation reps$.^{\cite{S}}$ It follows a spin  $(A,B)$ field must be accompanied by at least one spin $(C,D)$ field to have a nontrivial rep.

For example a Dirac spinor has composition $(1/2,0)\oplus (0,1/2).$ Thus the transformations of a Dirac spinor qualify and there is a nontrivial rep of translations along with rotations and boosts. One hint is the existence of Dirac gamma matrices, which are vector matrices. Vector matrices that commute are momentum matrices. In fact a Dirac spinor can realize two distinct reps of the Poincar\'{e} group. 

We proceed now to the derivation of the coefficients $u$ and $v,$ adapted from Ref. \cite{fields1}.  The coefficients $u$ and $v$ are determined by assuming ({\it{i}}) the operators transform by  $(\Lambda,b)$ as in (\ref{Da+}) , ({\it{ii}}) the coefficients are invariant,  and ({\it{iii}}) the fields transform by $(\Lambda,b)$  as in (\ref{Dpsi}).  By (\ref{psi+}), (\ref{Da+}) and (\ref{Dpsi}), and since the operators $a$ and $a^{\dagger}$ are linearly independent, the coefficients must obey
   $$ e^{i \Lambda p \cdot b} \sum_{\bar{l}} D_{l \bar{l}}(\Lambda,b) u_{\bar{l}\sigma}(x,{\overrightarrow{p}}) = \sqrt{\frac{ (\Lambda p)^t}{p^t}} \sum_{\bar{\sigma}} u_{l\bar{\sigma}}(\Lambda x + b,{\overrightarrow{\Lambda p}}) D^{(j)}_{\bar{\sigma} \sigma}(W(\Lambda,p))    \, $$
\begin{equation} \label{Du1}
   e^{-i \Lambda p \cdot b} \sum_{\bar{l}} D_{l \bar{l}}(\Lambda,b) v_{\bar{l}\sigma}(x,{\overrightarrow{p}}) = \sqrt{\frac{ (\Lambda p)^t}{p^t}} \sum_{\bar{\sigma}} v_{l\bar{\sigma}}(\Lambda x + b,{\overrightarrow{\Lambda p}}) {D^{(j)}}^{\ast}_{\bar{\sigma} \sigma}(W(\Lambda,p))    \, .
\end{equation}
To solve these equations for the coefficients we consider special cases.

As a first case, let $\Lambda$ = 1. By (\ref{WLp}), $W$ = $L^{-1}(p) L(p)$ = 1. Replacing  $x \rightarrow y,$ $b \rightarrow \bar{b},$ $p \rightarrow q$ in (\ref{Du1}), we have
$$   u_{l\sigma}( y ,{\overrightarrow{ q}}) = e^{-i q \cdot \bar{b}} \sum_{\bar{l}} D_{l \bar{l}}(1,-\bar{b}) u_{\bar{l}\sigma}(y+\bar{b},{\overrightarrow{q}})    \, $$
\begin{equation} \label{Du2}
   v_{l\sigma}( y ,{\overrightarrow{ q}}) = e^{i q \cdot \bar{b}} \sum_{\bar{l}} D_{l \bar{l}}(1,-\bar{b}) v_{\bar{l}\sigma}(y+\bar{b},{\overrightarrow{q}})    \, . 
\end{equation}

We use this case to relate the coefficients to those at the origin $x$ = 0. With $\bar{b}$ = $-y$ in (\ref{Du2}), we have
   $$ u_{l \sigma}( y,{\overrightarrow{q}}) = 
e^{i q \cdot y} \sum_{\bar{l}} D_{l \bar{l}}(1,y) u_{\bar{l}\sigma}(0,{\overrightarrow{q}})   \, $$
\begin{equation} \label{Du1a}
   v_{l \sigma}( y,{\overrightarrow{q}}) = 
e^{-i q \cdot y} \sum_{\bar{l}} D_{l \bar{l}}(1,y) v_{\bar{l}\sigma}(0,{\overrightarrow{q}})  \, .
\end{equation}
Next, consider (\ref{Du2}) for the subcases (a) $y$ = $x,$  $\bar{b}$ = $-x,$ $q$ = $p$ and (b) $y$ = $ \Lambda x + b,$ $\bar{b}$ = $-(\Lambda x + b),$ $q$ = $\Lambda p.$ Then the $u$s and $v$s in (\ref{Du1}) are dependent upon $u$s and $v$s at the origin and we have
 $$   \sum_{\bar{l}} D_{l \bar{l}}(\Lambda,0) u_{\bar{l}\sigma}(0,{\overrightarrow{p}}) =  \sqrt{\frac{ (\Lambda p)^t}{p^t}} \sum_{\bar{\sigma}} u_{l\bar{\sigma}}(0,{\overrightarrow{\Lambda p}}) D^{(j)}_{\bar{\sigma} \sigma}(W(\Lambda,p))    \, $$
\begin{equation} \label{Du4}\sum_{\bar{l}} D_{l \bar{l}}(\Lambda,0) v_{\bar{l}\sigma}(0,{\overrightarrow{p}}) = 
      \sqrt{\frac{ (\Lambda p)^t}{p^t}} \sum_{\bar{\sigma}} v_{l\bar{\sigma}}(0,{\overrightarrow{\Lambda p}}) {D^{(j)}}^{\ast}_{\bar{\sigma} \sigma}(W(\Lambda,p))   
\end{equation}

As a second case, let $\Lambda$ = $R$ be a rotation and  $p$ = $k$ = $\{0,0,0,m\}.$ Hence $\Lambda p$ = $Rk$ = $k.$ Since the standard transformation $L(p)$ here transforms $k$ to $p$ = $k,$ one infers that $L$ = $L^{-1}$ = 1 and, by (\ref{WLp}), $W$ = $R.$ By (\ref{Du4}), we have
 $$   \sum_{\bar{l}} D_{l \bar{l}}(R,0) u_{\bar{l}\sigma}(0,{\overrightarrow{0}}) = \sum_{\bar{\sigma}} u_{l\bar{\sigma}}(0,{\overrightarrow{0}}) D^{(j)}_{\bar{\sigma} \sigma}(R)    \, $$
\begin{equation} \label{Du5}
     \sum_{\bar{l}} D_{l \bar{l}}(R,0) v_{\bar{l}\sigma}(0,{\overrightarrow{0}}) = \sum_{\bar{\sigma}} v_{l\bar{\sigma}}(0,{\overrightarrow{0}}) {D^{(j)}}^{\ast}_{\bar{\sigma} \sigma}(R)    \, .
\end{equation}
By Schur's Lemma$,^{\cite{H1}}$ either the reps $D(R,0)$ and $ D^{(j)}(R)$ of the group of rotations $R$ are compatible or the coefficients vanish. 

By (\ref{Du5}) when the spin composition of the representation matrices $D$ is spin $(A,B)\oplus (C,D) \ldots ,$ the coefficients $u(0,{\overrightarrow{0}})$ and $v(0,{\overrightarrow{0}})$ are Clebsch-Gordan coefficients$,^{\cite{fields1}}$
  $$  u_{l \sigma}(0,{\overrightarrow{0}}) =  \frac{(2 \pi)^{-3/2}}{\sqrt{2m}} \pmatrix{  \langle AaBb \mid j \sigma \rangle \cr  \langle C c D d \mid j \sigma \rangle \cr \vdots   }     \, $$
 \begin{equation} \label{Du6}
    v_{l \sigma}(0,{\overrightarrow{0}}) =  \frac{(-1)^{j+\sigma}(2 \pi)^{-3/2}}{\sqrt{2m}} \pmatrix{  \langle AaBb \mid j, -\sigma \rangle \cr  \langle C c D d \mid j, -\sigma \rangle \cr \vdots   }     \, ,
\end{equation}
where $\langle AaBb \mid j \sigma \rangle$ is the Clebsch-Gordan coefficient which can be nonzero when spin $j \in$ $\{\mid A-B\mid, \ldots, A+B\}$ and $a+b$ = $\sigma.$ The factor $1/{\sqrt{2m}}$ is conventional and the factor $(2 \pi)^{-3/2}$ is included so that the $u$s and $v$s here match those in Ref. \cite{fields1}. Recall that the index $l$ on the left stands for the sequence of double indices $ab,cd, \ldots$ on the right.

As a third case, let  $\Lambda$ be the special transformation $L(q)$ taking $k$ to $q$ and let $p$ = $k$ = $\{0,0,0,m\}.$ Then, by (\ref{WLp}), $W$ = $L^{-1}(q) L(q)$ = 1 and, by (\ref{Du4}), one finds that
$$   u_{l{\sigma}}(0,{\overrightarrow{q}}) = \sqrt{\frac{ m}{q^{\, t}}} \sum_{\bar{l}} D_{l \bar{l}}(L,0) u_{\bar{l}\sigma}(0,{\overrightarrow{0}})     \,$$ 
\begin{equation} \label{Du7}
   v_{l{\sigma}}(0,{\overrightarrow{q}}) = \sqrt{\frac{ m}{q^{\, t}}} \sum_{\bar{l}} D_{l \bar{l}}(L,0) v_{\bar{l}\sigma}(0,{\overrightarrow{0}})     \, ,
\end{equation}
where $L$ = $L(q).$ By (\ref{Du2}), (\ref{Du6}) and (\ref{Du7}), one finds an expression for the coefficients $u$ and $v,$
 %\pagebreak
$$   u_{l\sigma}( x ,{\overrightarrow{p}}) = (2 \pi)^{-3/2}\sqrt{\frac{ 1}{2p^{\, t}}} \, e^{i p \cdot x} \sum_{\bar{l}} D_{l \bar{l}}(L,x) \pmatrix{  \langle A\bar{a}B\bar{b} \mid j \sigma \rangle \cr  \langle C \bar{c} D \bar{d} \mid j \sigma \rangle \cr \vdots   }     \, ,$$
\begin{equation} \label{Du8}
    v_{l\sigma}( x ,{\overrightarrow{p}}) =  (-1)^{j+\sigma} (2 \pi)^{-3/2}\sqrt{\frac{ 1}{2p^{\, t}}} \, e^{-i p \cdot x} \sum_{\bar{l}} D_{l \bar{l}}(L,x) \pmatrix{  \langle A\bar{a}B\bar{b} \mid j, -\sigma \rangle \cr  \langle C \bar{c} D \bar{d} \mid j, -\sigma \rangle \cr \vdots   }     \, ,
\end{equation}
where $L$ = $L(p)$ is the special transformation taking $k$ = $\{0,0,0,m\}$ to $p$ and the single index $\bar{l}$ on $D_{l\bar{l}}$ is the double index $\{\bar{a} \bar{b},\bar{c} \bar{d},\ldots \}$ in the column matrix of Clebsch-Gordan coefficients.

The expressions (\ref{Du8}) for the coefficients $u$ and $v$ along with expressions (\ref{psi+}) for $\psi^{+}$ and $\psi^{-}$ determine the field $\psi_{l}(x)$ = $\kappa \psi^{+}_{l}(x)$ + $\lambda \psi^{-}_{l}(x).$

\section{Discussion } \label{discuss}

By (\ref{Du8}), the coefficients $u$ and $v$ depend on the matrix $D(L,x)$ representing the spacetime transformation $(L(p),x)$ which is the standard Lorentz transformation taking the rest momentum $k$ = $(0,0,0,m)$ to the momentum $p^{\mu}$ followed by a translation along the displacement $x.$ One has  
$$ D(L,x) = D(1,x)D(L,0) \, .$$
By (\ref{psi+}), to get the field $\psi,$ one integrates expressions linear in $u$ and $v$ over $\overrightarrow{p}$ with $x$ held constant. Since $L$ depends on $p,$ the matrix $D(L,0)$ cannot be factored out of the integrand, but $D(1,x)$ can. Moreover, the same $D(1,x)$ occurs with both the annihilation and the creation fields, and it follows, by linearity, that
 \begin{equation} \label{psi5}
    \psi_{l}(x) =  D_{l \bar{l}}(1,x) \psi_{\bar{l}}^{(0)}(x)   \, ,
\end{equation}
where $\psi_{\bar{l}}^{(0)}(x)$ is the field obtained with the homogeneous Lorentz rep$.^{\cite{fields1,fields2}}$

Consider a field $\psi$ with spin $(A,B) \oplus (C,D).$ Assume the angular momentum and boost matrices are block diagonal with spin $(A,B)$ in the 11 block and spin $(C,D)$ in the 22 block. Then translations are generated by momentum matrices in one of two forms, one with the off-diagonal 12 block of the $P^{\mu}$ nonzero and one with the 21 block nonzero. For details see Ref. \cite{S}. 

Consider the 12-representation; the discussion for the 21 representation is much the same. Translation through a spacetime displacement of $x^{\mu}$ is represented by the matrix
\begin{equation} \label{T} D(1,x^{\mu})  = \exp{(-i \eta_{\mu \nu} x^{\mu}P^{\nu})} =  \pmatrix{1 && -i x_{\mu}P^{\mu}_{12} \cr 0 && 1} \quad ,\end{equation}
where $x_{\mu}$ = $\eta_{\mu \nu} x^{\nu},$ and the metric $\eta$ is diagonal; we use $\eta$ = diag$\{1,1,1,-1\}.$ The matrix exponential simplifies because only the 12-block is nonzero, making quadratic and higher orders vanish,
\begin{equation} \label{block} P^{\mu}P^{\nu} =  \pmatrix{0 && P^{\mu}_{12} \cr 0 && 0} \pmatrix{0 && P^{\nu}_{12} \cr 0 && 0} = \pmatrix{0 && 0 \cr 0 && 0}  \quad . \end{equation}
By (\ref{psi5}), applying the translation $D(1,x)$ to field $\psi^{(0)}(x)$ yields the field $\psi(x),$
\begin{equation} \label{T1} \psi(x) = D(1,x^{\mu})\pmatrix{\psi^{(0)}_{ab}(x) \cr \psi^{(0)}_{cd}(x)} =  \pmatrix{\psi^{(0)}_{ab}(x) +\lambda_{ab} \cr \psi^{(0)}_{cd}(x)} \, , \end{equation}
where
$$\lambda_{ab} = -ix_{\mu} \sum_{\bar{c}\bar{d}} P^{\mu}_{12;ab,\bar{c}\bar{d}} \psi^{(0)}_{\bar{c}\bar{d}}(x) \, .$$
The $CD$ block of $\psi(x)$ is unchanged from the $CD$ block of $\psi^{(0)}(x).$ 

We see that the translation $D(1,x)$ adds an inhomogeneous term to the $AB$ block $\psi^{(0)}_{ab}(x),$ $\psi^{(0)}_{ab}(x) \rightarrow $ $\psi^{(0)}_{ab}(x) + \lambda_{ab}.$ This is a connection-like transformation typical of translations$^{\cite{M}}$ and unlike boosts and rotations which are homogeneous transformations. 

The scale of the inhomogeneous term $\lambda$ in (\ref{T1}) is not determined since the scale of the momentum matrices, let's call it $k_{(12)},$ is a free parameter. One can see this from the commutation rules of the Poincar\'{e} algebra that involve the momentum $P$: $[J_x,P_y]$ = $i P_z,$  $[K_x,P_x]$ =  $-i  P_t,$ $[K_x,P_y]$ = 0, $[K_x,P_t]$ =  $-i P_x,$ and $[P_\mu,P_\nu]$ = 0, and permutations thereof. Thus the distance or time scale, $x^{\mu} \approx$ $1/k_{(12)},$ at which, in general, $\lambda$ becomes comparable to $\psi^{(0)}_{ab},$ remains undetermined.

%\pagebreak

\appendix

%\pagebreak

\section{Problems} \label{B}

\vspace{0.3cm}
\noindent 1. (a) Write $D(\Lambda,b)$ as a translation followed by a homogeneous Lorentz transformation. That is, find $\bar{\Lambda}$ and $\bar{b}$ in $D(\Lambda,b)$ =  $D(\bar{\Lambda},0)$$D(1,\bar{b}).$ 
 
\vspace{0.3cm}
(b) Find expressions for $D^{(-1)}(\Lambda,b)$ (i) as  a Lorentz transformation followed by a translation and (ii) as a translation followed by a Lorentz transformation.

\vspace{0.3cm}
\noindent 2. In order that spin $(A,B) \oplus (C,D)$ have nontrivial translation matrices, what spins $(C,D)$ can accompany 
 
\vspace{0.3cm}
(a) $(A,B)$ = $(0,1/2)$  ?
 
\vspace{0.3cm}
(b) $(A,B)$ = $(1/2,1/2)$ ? 
 
\vspace{0.3cm}
(c) What particle spins $j$ are allowed for each $(A,B) \oplus (C,D)$ found in (a) and (b)  (i) when neither $\psi^{(0)}_{ab}(x) $ nor $ \psi^{(0)}_{cd}(x)$ is identically zero and (ii) when just one of the fields $\psi^{(0)}_{ab}(x) $ and $ \psi^{(0)}_{cd}(x)$ vanishes?

\vspace{0.3cm}
\noindent 3. Since $p^2$ = $-M^2,$ the field ${\psi^{(0)}}(x)$ obeys the differential equation $(\eta^{\mu \nu} \partial_{\mu} \partial_{\nu} - M^2){\psi^{(0)}}(x)$ = 0. Find the corresponding differential equation satisfied by  $\psi(x).$ 

\vspace{0.3cm}
\noindent 4. For spin $(A,B) \oplus (C,D)$ = $(0,1/2) \oplus (1/2,0),$ it can be shown that the field $\psi^{(0)}(x)$ in (\ref{psi5}) satisfies the Dirac equation. See Ref. \cite{W7} for details. Now suppose that $\psi^{(0)}(x)$ satisfies the Dirac equation, $(\gamma^{\mu} \partial_{\mu} + M)\psi^{(0)}(x)$ = 0, where $\gamma^{\mu}$ satisfy
$$\gamma^{\mu}. \gamma^{\nu} + 
\gamma^{\nu}. \gamma^{\mu} = 2 \eta^{\mu \nu} I \quad ,$$ where $I$ is the $4\times 4$ unit matrix.

\vspace{0.3cm}
(a) Obtain angular momentum matrices $J^{\mu \nu}$ by using $$\gamma^{\mu}. \gamma^{\nu} - 
\gamma^{\nu}. \gamma^{\mu} = 4 i J^{\mu \nu} \quad .$$ Note that $J^{12}$ = $J^{z}$ and $J^{14}$ = $K^{x},$ etc. Show that the $J^i$ and $K^i$ matrices satisfy the Lorentz commutation rules $[J^x,J^{y}]$ = $i J^{z},$ $[J^x,K^{y}]$ = $i K^{z},$ and $[K^x,K^{y}]$ = $-i J^{z},$ along with cyclic permutations of $\{x,y,z\}.$

 \vspace{0.3cm}
(b) Show that the gamma matrices are vector matrices by showing that they satisfy the commutation rules $[J^x,\gamma^{y}]$ = $i \gamma^{z},$ $[J^x,\gamma^{t}]$ = 0, $[K^i,\gamma^{j}]$ = $i \gamma^{t}\delta^{ij},$ and $[K^x,\gamma^{t}]$ = $i \gamma^{x},$ with permutations.

 \vspace{0.3cm}
(c) Let $\gamma_{5}$ = $-i\gamma^{t}\gamma^{x}\gamma^{y}\gamma^{z},$ $P^{\mu}_{12}$ = $(1+\gamma_{5})\gamma^{\mu}/2,$ and $P^{\mu}_{21}$ = $(1-\gamma_{5})\gamma^{\mu}/2.$ Show that the matrices $P^{\mu}_{12}$ satisfy the commutation rules in (c) and show that they commute amongst themselves, $[P_{12}^{\mu},P_{12}^{\nu}]$ = 0. Thus the matrices $P_{12}^{\mu}$ are a set of momentum matrices and the set $\{J^{i},K^{j},P_{12}^{\mu}\}$ generates a representation of the Poincar\'{e} group of rotations, boosts and translations. Likewise for the $P_{21}^{\mu}.$

\vspace{0.3cm}
(d) What equation does $\psi(x)$ satisfy with the $\{J^{i},K^{j},P_{12}^{\mu}\}$ rep? with the $\{J^{i},K^{j},P_{21}^{\mu}\}$ rep?

\vspace{0.3cm}
\noindent 5. Continue with spin 1/2 as in Problem 4. Let $\phi^{(0)}(t)$ be a plane wave, $\phi^{(0)}(t)$ = $e^{-i mt} \phi,$ where $\phi$ is an eigenvector of $\gamma^{t},$ with $v^{(0)\mu }$ = $ \bar{\phi}^{(0)}(t) \cdot \gamma^{\mu} \cdot \phi^{(0)}(t) $  = $\{0,0,0,-1\},$ where $\bar{\phi}^{(0)}(t) $ = ${\phi^{(0)}}^{\dagger}(t) \cdot \gamma^{t}.$ Conventionally, $\phi^{(0)}(t)$ describes a stationary particle. Let $\phi(t)$ be the generalized wave $\phi(t)$ = $D(1,t) \phi^{(0)}(t),$ where $D(1,t)$ is the translation matrix for duration $t$ with no spatial displacements, $x^x$ = $x^y$ = $x^z$ = 0. Define $v^{\mu}(t)$ = $ \bar{\phi}(t).\gamma^{\mu}.\phi(t).$ Find an eigenvector $\phi$ such that $v^{\mu}(t) - v^{(0)\mu }$ is light-like and (i) in the $xt$-plane and (ii) in the $zt$-plane.

\end{document}